\documentclass{emulateapj}
\usepackage{graphicx}
\usepackage{layout}
\usepackage[latin1]{inputenc}
\usepackage{amssymb}
\usepackage{natbib}

\slugcomment{To appear in ApJ}
\shorttitle{The nature of SDSS J1536+0441}
\shortauthors{Decarli et al.}

\def\lsim{\mathrel{\rlap{\lower 3pt \hbox{$\sim$}} \raise 2.0pt \hbox{$<$}}}
\def\gsim{\mathrel{\rlap{\lower 3pt \hbox{$\sim$}} \raise 2.0pt \hbox{$>$}}}

\shorttitle{Probing the nature of SDSS J1536+0441}
\shortauthors{Decarli et al.}

\begin{document}

\title{
Probing the nature of the massive black hole binary candidate SDSS J1536+0441\footnotemark[1]\footnotetext[1]{Based on observations collected at the European Organisation for Astronomical Research in the Southern Hemisphere, Chile (DDT programme: 282.B-5059).}
}

\author{R. Decarli\altaffilmark{2}, M. Dotti\altaffilmark{3}, R. Falomo\altaffilmark{4}, 
A. Treves\altaffilmark{2}, M. Colpi\altaffilmark{5}, J.K. Kotilainen\altaffilmark{6}, 
C. Montuori\altaffilmark{5}, M. Uslenghi\altaffilmark{7}}
\altaffiltext{2}{Dipartimento di Fisica e Matematica, Universit\`a dell'Insubria,
via Valleggio 11, I-22100 Como, Italy. E-mail: {\sf roberto.decarli@mib.infn.it}}
\altaffiltext{3}{Department of Astronomy, University of Michigan, Ann Arbor, MI 48109, USA}
\altaffiltext{4}{INAF -- Osservatorio Astronomico di Padova, Vicolo dell'Osservatorio 5,
I-35122 Padova, Italy}
\altaffiltext{5}{Dipartimento di Fisica, Universit\`a degli Studi di Milano--Bicocca, piazza delle Scienze 3, 20126 Milano, Italy}
\altaffiltext{6}{Tuorla Observatory, Department of Physics and Astronomy,
University of Turku, V\"ais\"al\"antie 20, FI--21500 Piikki\"o,
Finland}
\altaffiltext{7}{INAF-IASF Milano, Via E. Bassini 15, Milano I-20133, Italy
}

\begin{abstract}
We present an imaging study of the black hole binary candidate
SDSS J1536+0441 ($z\approx0.3893$), based on deep, high resolution 
V$z$K images collected at the ESO/VLT. 
The images clearly show an asymmetric elongation, indicating the
presence of a companion source at $\sim 1''$ ($\approx5$ kpc projected
distance) East from the quasar. The host galaxy of the quasar is 
marginally resolved. We find that the companion source is a 
luminous galaxy, the light profile of which suggests the presence of an 
unresolved, faint nucleus (either an obscured AGN or a compact stellar
bulge). The study of the environment around the quasar indicates the
occurrence of a significant over-density of galaxies with a redshift 
compatible with $z\approx0.4$. This suggests that it resides in 
a moderately rich cluster of galaxies.
\end{abstract}

\section{Introduction}

\citet{boroson09} discovered peculiar features in the optical
spectrum of \object{SDSS J153636.22+044127.0} (hereon, SDSS
J1536+0441), a $z=0.3893$ radio quiet quasar. Three line systems are
found at different redshifts: a set of broad and narrow emission
lines at $z=0.3893$, a set of only broad emission lines at
$z=0.3727$ and a third set of very narrow absorption lines at an
intermediate redshift ($z=0.3878$). \citet{boroson09} interpreted
the former two systems in terms of the broad line region emission
around two massive black holes within the same galaxy, constituting
a massive black hole binary (BHB) with sub-parsec separation.

\citet{chornock09a,chornock09b} questioned the BHB scenario,
ascribing the peculiar spectroscopic features of the quasar to a
single active massive black hole with a particular disk-like
geometry of the broad line region, similar to Arp 102B. This
scenario is also supported by the detection of a bump in the red
wings of broad emission line profiles, at $z\approx0.4$.

Motivated by the claim by \citet{boroson09}, \citet{wrobel09}
observed SDSS J1536+0441 using the  VLA, discovering two radio
sources (VLA--A  and VLA--B)  separated by $0.97''$ within the
quasar  optical localization  region. Independently, NIR
observations led by our group \citep{decarli09b} revealed the
occurrence of a companion galaxy of the quasar, located in
correspondence with VLA--B. This was subsequently confirmed in the
optical wavelengths by \citet{lauer09} using HST.

\citet{wrobel09} and \citet{decarli09b} suggested that each nucleus
of the two galaxies is related to a single set of broad emission
lines. The validity of this picture can be tested with high angular
resolution spectroscopy, in order to prove if the two sets of broad
emission lines are spatially separated. \citet{lauer09} and
\citet{chornock09b} presented optical spectra of SDSS J1536+0441.
Both groups suggest that the regions responsible for the emission of
the two sets of broad lines are spatially coincident. Nevertheless,
both observations were carried out under non-optimal seeing
conditions (larger than the angular separation of the two objects
along the slit), hindering a clear comprehension of the contribution
of the active nucleus in VLA--B to the spectrum of SDSS J1536+0441.

The scenario of a superposition of two AGNs has already been invoked by
\citet{heckman09} to explain the spectral properties of another BHB
candidate \citep[SDSS J0927+2943;
see][]{komossa08,dotti09,bogdanovic09}, but was subsequently
questioned by \citet{decarli09a}. \citet{boroson09} noted that the
expected number of chance quasar pairs with projected separations
$\lsim1''$ in the whole SDSS DR7 sample is only 0.3\%. On the other
hand, if the two objects are physically linked (e.g., if they lie in
a rich galaxy cluster), such a number rises significantly \citep[see
the discussion in][]{wrobel09}.

In order to study the properties of this peculiar object, and to
probe the presence of a rich galaxy cluster in its environment, we
performed deep, high-resolution imaging in the optical and near
infrared wavelengths in an ESO Director Discretional Time programme
(proposal ID: 282.B-5059).

Throughout the paper, we adopt a concordance cosmology
with $H_0=70$ km/s/Mpc, $\Omega_{\rm m}=0.3$, $\Omega_\Lambda=0.7$.

\section{Observations, data reduction and analysis}

\begin{figure}
\begin{center}
\includegraphics[width=0.49\textwidth]{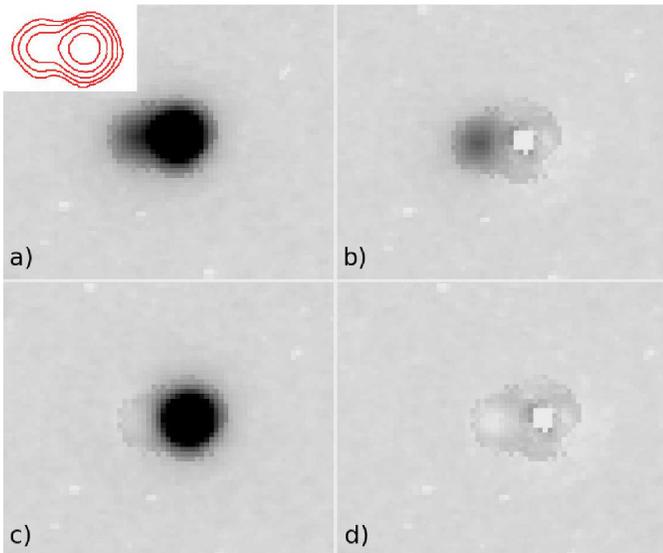}\\
\caption{K-band images of SDSS J1536+0441. Each panel is $7''\times6''$
wide. North is up, East to the left. {\bf a)} The quasar image, as
observed in the K-band. The companion galaxy corresponding to VLA--B
makes the quasar image clearly oblate towards the left. The
contour plots of the radio detection by \citet{wrobel09} are also
shown in the inset, plotted with the same size scale.
{\bf b)} Image of the companion source, after the subtraction of the 
quasar model.
{\bf c)} Image of the quasar, after the subtraction of the model
of source B. {\bf d)} Residuals after the subtraction of both the
models. No obvious asymmetry in the residual map of the 
two sources is reported, suggesting that the two galaxies are 
fairly axisymmetric.}\label{fig_q1536K}
\end{center}
\end{figure}

K-band observations of the field around SDSS J1536+0441 were
collected with the High Acuity Wide field K-band Imager (HAWK-I) at
the ESO/Very Large Telescope (VLT) on April, 30, 2009. The total
integration time amounts to 48 minutes. HAWK-I is equipped with 4
CCDs of 2048$\times$2048 pixels and a pixel-scale of $0.1064''/$pxl.
Each CCD image $3.2\times3.2$ square arcmin (after trimming and the
removal of partly observed regions). Usual jitter procedure was
applied in order to collect broad band images in the NIR. Images
were reduced with an IRAF\footnotemark[7]\footnotetext[7]{IRAF is
distributed by the National Optical Astronomy Observatories, which
are operated by the Association of Universities for Research in
Astronomy, Inc., under cooperative agreement with the National
Science Foundation}-based pipeline developed and tested in other
programmes \citep[e.g.,][]{kotilainen09}.
The seeing was $0.75-0.80''$ (FWHM) throughout the whole 
integration time.
We estimate the Zero Point by comparing the instrumental magnitudes
of field sources with the values reported in the 2MASS database. The
accuracy in the photometric calibration is $0.03-0.05$ mag.
Hereafter, we will focus on the Chip 2 image, the one overlapping
onto the optical observations.

V- and $z$-band images were collected with the Focal Reducer and low
dispersion Spectrograph \#2 (FORS2) at the ESO/VLT on May, 22 and
24, 2009. FORS2 mounts 2 CCDs of 2048$\times$4096 pixels. Our
observations were carried out with the High-Resolution collimator
with $1\times 1$ binning, yielding a pixel-scale of $0.0632''/$pxl.
Offsets within a $10''$ wide box were applied in order to overcome
the $3.5''$ gap between the two FORS2 CCDs. The final frames are
$3.6\times3.3$ square arcmin wide. In the V-band, ten, 90 sec-long
images were acquired, for a total integration time of 15 minutes. In
the $z$-band, several 120-sec long exposures were secured, yielding
a total integration time of 52 minutes. The seeing was excellent
(FWHM=$0.7''$ in $z$, $0.4''$ in V). The Zero Point of the $z$-band
observation was estimated from the $z$ magnitudes of field sources
in the SDSS photometric catalogue. Similarly, for the V-band, we
referred to the V magnitude estimates as derived from the SDSS $g$
and $r$ bands, following the conversion recipes by
\citet{windhorst91}. The internal accuracy is around $0.04$ mag.

A composite image of our data in the three observed bands is available 
on-line at \verb|www.dfm.uninsubria.it/astro/qso_host/q1536/|.

The Astronomical Image Decomposition and Analysis
\citep[AIDA,][]{uslenghi08} software is used to model the Point
Spread Function (PSF). The employed technique is widely discussed in
\citet{kotilainen07,kotilainen09}, and it is briefly summarized
here. The images of field stars are modeled with the superposition
of 4 bidimensional gaussians mimicking the core plus an exponential
feature for the wings of the PSF. No significant PSF variation
through the frame is reported in the K-band frame. On the other
hand, FORS2 images are affected by severe distortions of the PSF
shape throughout the image, making the PSF model less secure.

\section[]{Results}\label{sec_results}

\subsection{Decomposition of the quasar image}

\begin{figure}
\begin{center}
\includegraphics[width=0.49\textwidth]{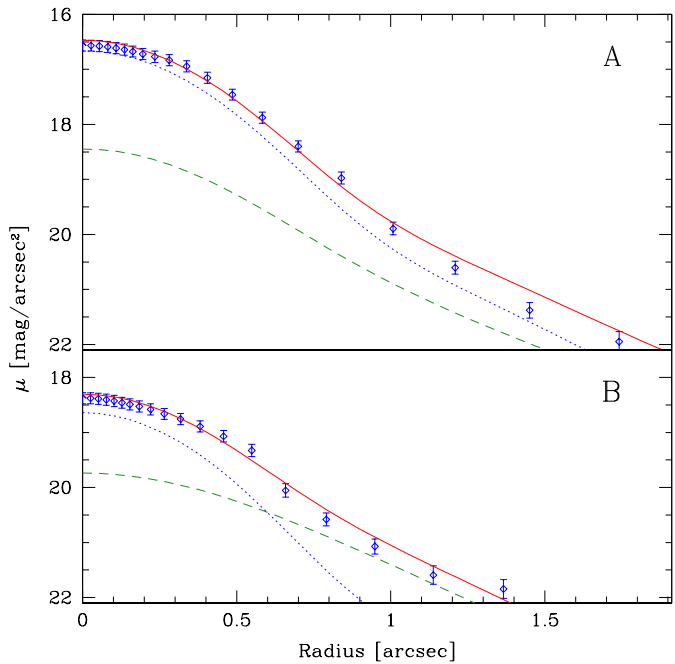}\\
\caption{The K-band light profiles of SDSS J1536+0441 (A) and its
companion source (B) as compared to the best fit model (solid line),
the PSF model (dotted line) and the galaxy model (dashed line).
The light profile of quasar A shows deviations from the
model at large radii, suggesting that the host galaxy may present stellar
shells \citep[cfr][]{canalizo07}.
}\label{fig_resolved}
\end{center}
\end{figure}

Figure \ref{fig_q1536K} shows that the image of SDSS J1536+0441 is
clearly elongated East-wards, revealing the presence of a blended
source (hereafter, source B) close to the quasar and coincident with
VLA--B. In a comparison between the PSF model and the observed light
profile of the two sources, it is apparent that they are both
resolved (see Figure \ref{fig_resolved}). As quasar A is blended
with its faint companion, in order to study the presence of extended
emission in both of them we adopt an iterative procedure to model
the two sources: 1) First, we mask the Eastern side of the bright
quasar (where the contribution from the companion source may be
relevant) and perform a 2D modelling with a nuclear+host galaxy
profile convolved to our PSF model. 2) The model is subtracted from
the observed frame, the residuals are masked out and we model the
emission from the companion source. 3) The companion model is
subtracted from the original observed frame, and we re-model the
emission from the quasar. 4) Finally, we re-perform the second step
with the new model of the main source. 

\begin{table*}
\caption{{\rm Best fit parameters in the modelling of the image of
SDSS J1536+0441. (1) Band of the observations. (2) Total apparent
magnitude of the model. (3) Apparent magnitude of the unresolved
component. (4) Apparent magnitude of the host galaxy model. (5)
Comparison between the $\chi^2$ obtained assuming a pure PSF and a
galaxy+nucleus model. (6) The same as 5, comparing the cases without
and with a nuclear component. (7) Rest frame band. (8) Absolute
magnitude of the nucleus in the galaxy+nucleus decomposition. (9) The
same as 8, for the galaxy. }} \label{tab_fitres}   
\begin{center}

\begin{tabular}{cccccc|ccc}
   \hline
   \hline
   Band & $m_{\rm tot}$ & $m_{\rm nuc}$ & $m_{\rm gal}$ &$\chi_{\rm nuc}^2/\chi_{\rm gal+nuc}^2$& $\chi_{\rm gal}^2/\chi_{\rm gal+nuc}^2$ & RF-Band & $M_{\rm nuc}$ &  $M_{\rm gal}$\\
    (1) &  (2)      &  (3)      &  (4)    &(5)   & (6)  &  (7) & (8)   & (9)   \\
   \hline                                 %
   \hline
   \multicolumn{8}{l}{\emph{Source A}} \\              %
   \hline
     K  &  14.1     & $14.5\pm0.3$ & $15.6\pm0.6$ & 1.6  & 3.7  & H & $-26.2\pm0.3$ & $-25.3\pm0.6$ \\ 
    $z$ &  16.2     & $16.4\pm0.2$ & $\sim18$	  & 1.0  & 3.2  & R & $-24.8\pm0.2$ & $\sim-23$     \\ 
     V  &  17.5     & $17.6\pm0.2$ & $\sim21$	  & 1.0  & 4.4  & B & $-23.7\pm0.2$ & $\sim-21$     \\ 
  \hline							     
  \hline				        		     
   \multicolumn{8}{l}{\emph{Source B}} \\       		     
  \hline				       
     K  &  15.8     & $16.6\pm0.3$ & $16.5\pm0.3$ & 2.6  & 1.9  & H & $-24.2\pm0.3$ & $-24.6\pm0.3$ \\ 
    $z$ &  18.6     & $20.0\pm0.3$ & $18.9\pm0.3$ & 4.9  & 1.4  & R & $-21.3\pm0.3$ & $-22.4\pm0.3$ \\ 
     V  &  21.0     & $23.5\pm0.4$ & $21.2\pm0.6$ & 3.6  & 1.3  & B & $-19.0\pm0.4$ & $-20.6\pm0.6$ \\ 
  \hline
  \hline
\end{tabular}
   \end{center}
\end{table*}

In order to evaluate the uncertainties in the magnitudes of nuclear and
extended emissions, we scaled the Nuclear-to-Host luminosity ratio for
a large set of models, and conservatively adopted as reference those 
bracketing the observed light profiles. 
Table \ref{tab_fitres} summarizes the results of the fit procedure.
The host galaxy of quasar A is significantly detected only in the
K-band. On the contrary, source B is well resolved. In all the
bands, adding an unresolved nuclear component significantly improves
the quality of the fit. As the unresolved-to-resolved luminosity
ratio in the $z$ band ($\approx$ rest-frame R band, see below) is
around $0.3$, we cannot assess whether it is a low-luminosity active
nucleus (as suggested by the radio detection) or a stellar bulge.

The observed V, $z$, K bands correspond to the rest-frame B, R and
H. We compute the absolute magnitudes in these bands as
$M_{x}=m_{y}-DM-A_{y}+C_{x-y}$, where $x$ refers to the rest-frame
bands (B,R,H), $y$ the observed ones (V,$z$,K), $DM$ is the distance
modulus, $A_{y}$ is the Galactic extinction and $C_{x-y}$ is a term
accounting for filter and $k$-correction. In our cosmological
framework, $z=0.3893$ yields $DM=41.62$. The Galactic extinction is
derived from the HI maps by \citet{schlegel98}. The $k$-correction
comes from the galaxy templates by \citet{mannucci01} and the
quasar template by \citet{vandenberk01}.
The resulting magnitudes are reported in table \ref{tab_fitres}.
Source A has host galaxy absolute magnitudes consistent with the
typical quasar host galaxy luminosities \citep[e.g.,][]{kotilainen09} 
and it is $\sim2$ mag brighter than the
characteristic luminosity of quiescent galaxies at this redshift
\citep{cirasuolo08}. The companion source is a galaxy $\sim 1$ mag 
fainter \citep[cfr also the $M_g=-21.4$ estimate reported
by][]{lauer09} with a much fainter, red nucleus.

We note that, if the B-band flux of the unresolved source is
interpreted in terms of an active galactic nucleus, source B should
be classified as radio loud, according to the standard definition
\citep[e.g.,][]{kellermann89} based on the radio-to-optical flux
ratio \citep[here, $R\sim150$, assuming the radio flux found
by][]{wrobel09}.

\subsection{The properties of the environment}

In order to investigate the clustering properties of the environment
of this quasar, we search for galaxies in its field. We consider the
$3.2\times3.2$ square arcmin ($\approx 1\times1$ Mpc at $z=0.4$)
region that has been gathered in all the bands. We produce an
inventory of sources exceeding a 5-$\sigma$ detection threshold with
respect to the background count rms through the software SExtractor
\citep{bertin96}. The corresponding limit magnitudes are $m_{{\rm
V,}z{\rm ,K}}^{\rm lim}=25.7$, $23.1$ and $19.8$. We then match 
the compilations of sources detected in the
various bands. The complete list consists of 200 sources detected in
the V-band, 165 in $z$ (153 of which also appearing in the V-band
catalogue) and 87 in K (80 of which also detected in V). A total of
69 sources are detected in all the bands.

\begin{figure}
\begin{center}
\includegraphics[width=0.49\textwidth]{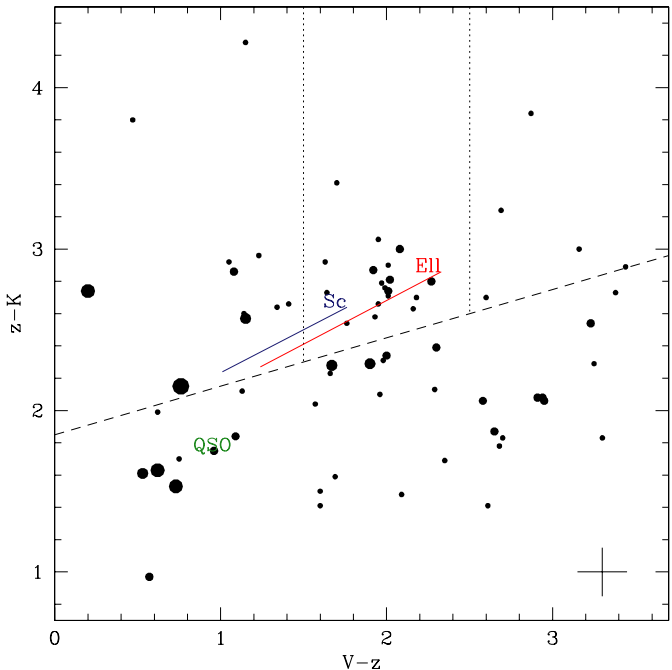}\\
\caption{The VzK color diagram for the objects in our analysis.
Symbol size maps the object brightness. `Ell', `Sc' and `QSO' label
the expectation from the elliptical and Sc galaxy templates by
\citet{mannucci01} and from the quasar template by \citet{vandenberk01}, 
redshifted to $z=0.4$. The solid lines show
how galaxy colors change from $z=0$ to $z=0.4$. The typical
error box is shown as a cross in the corner of the frame. The color cut in
equation \ref{eq_VzK} is shown as a dashed line. A number of
sources lie within the expectation for galaxies at $z\approx0.4$
(the vertical, dotted lines), suggesting the presence of a galaxy
cluster. }\label{fig_ccdiag}
\end{center}
\end{figure}

Among them, we select galaxy candidates using the color--color
diagram proposed by \citet{blanc08} and adapted to our photometric system
(see Figure \ref{fig_ccdiag}). We assume that a galaxy has:
\begin{equation}\label{eq_VzK}
(z-{\rm K})>0.3 ({\rm V}-z)+1.85
\end{equation}
Out of 69 sources, 34 match this color criterion.
We note that, given our sensitivity limits, we expect to detect $\sim30$
stars in our frame in all the bands at this Galactic
latitude\footnotemark[8]\footnotetext[8]{Based on the
Galaxy simulations performed with the software TRILEGAL available at
\texttt{http://stev.oapd.inaf.it/cgi-bin/trilegal} \citep{girardi05}.}.
Similarly, if we consider the number density of general field galaxies
\citep[e.g., from][]{cirasuolo08,blanc08}, 14 candidates brighter
than our sensitivity limits are expected. This indicates
that the field around the quasar is over-dense. From the galaxy
templates by \citet{mannucci01}, we infer that a $z\approx0.4$ galaxy has V-$z$
ranging between 1.5 and 2.5 \citep[see also][]{csabai03}; thus, 18 out
of 34 galaxy candidates are consistent with being at $z\approx0.4$.
The corresponding absolute magnitudes of these objects, assuming that 
they are galaxies at the same redshift as the quasar, are 
$\langle M_{\rm B}\rangle = -22.8$, $\langle M_{\rm R}\rangle = -20.5$ and 
$\langle M_{\rm H}\rangle = -19.2$ (with rms $\sim1$ mag), consistent with 
typical fairly bright galaxies \citep[cfr, e.g.,][]{gavazzi03}. It is
interesting to note that, out of 11 well-resolved sources in the K-band,
10 are classified as galaxies in our color selection, and 9 of them are 
consistent with having $z\approx0.4$.

\begin{figure}
\begin{center}
\includegraphics[width=0.49\textwidth]{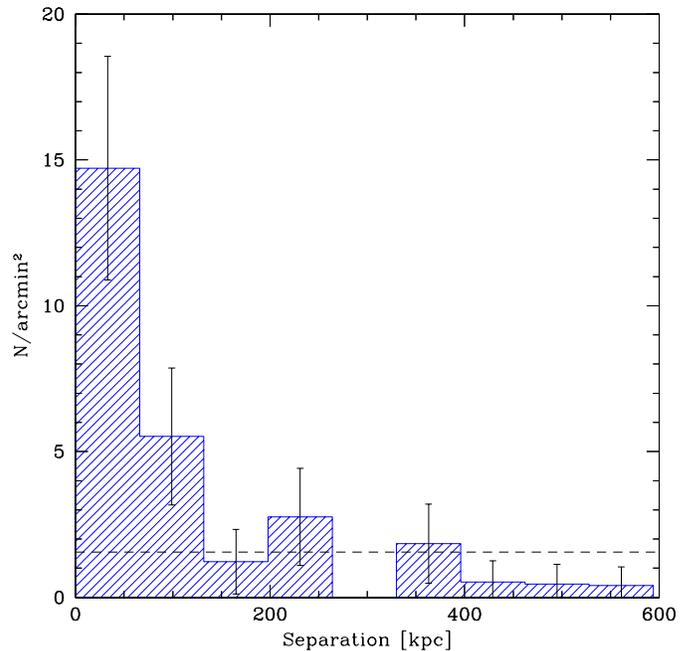}\\
\caption{The number density of the galaxy candidates consistent
with $z=0.4$ as a function of the projected distance
from SDSS J1536+0441. For a comparison, the expected number density
of background galaxies is plotted as a dashed line. A clear excess of 
galaxy candidates within $\sim 200$ kpc from the quasar is observed, 
supporting the cluster scenario.
}\label{fig_sd}
\end{center}
\end{figure}

We compute the number density of these galaxies as a function of the
angular distance from the quasar in order to evaluate the occurrence of
any clustering (see Figure \ref{fig_sd}). We find a significant excess of
the galaxy density within $\lsim200$ kpc from SDSS J1536+0441 with respect
to average field. Thus the quasar is found in a moderately rich cluster of 
galaxies.


\section{Discussion}\label{sec_conclusions}

In this Letter, we presented a high-resolution, deep multi-color 
imaging study of the alleged binary quasar SDSS J1536+0441. In addition 
to the extended emission of the host galaxy, we detect a bright companion
galaxy at $\sim 1''$ from the quasar. The quasar and the companion are 
located in correspondence with two distinct radio sources observed by 
\citet{wrobel09}, showing that both galaxies host an AGN.

Our deep images unveil the presence of a rich galaxy cluster. The
existence of such a massive structure solves the apparent inconsistency
between the observation of two radio sources within $\lsim1''$ and the
expected number of such pairs in the SDSS catalogue, as discussed in
\citet{boroson09} and \citet{wrobel09}. Furthermore, the potential well
of such a rich galaxy cluster could in principle explain the shift in
redshift between the two sets of emission lines. This consideration
favours a superposition picture to explain the spectrum of SDSS J1536+0441.

However, the red color and the faintness of the nuclear optical continuum 
observed in the companion galaxy suggest that its AGN emission is obscured. 
As a consequence, the AGN of the companion galaxy should not contribute 
much to the broad emission lines present in the SDSS spectrum, in 
agreement with the indications discussed in \citet{chornock09b} and 
\citet{lauer09}. In this case, the two sets of broad emission lines 
should be emitted by the quasar, consistently with the BHB or the double 
peaked emitter scenarios. The external regions of the companion 
galaxy could also naturally explain the narrow absorption lines observed 
in the spectrum of SDSS J1536+0441, as already suggested by \citet{lauer09}.
The origin of the red bump in the broad line profiles reported by 
\citet{chornock09b} remains unclear both in the BHB and the superposition 
scenarios.

The effective role of the companion in the spectrum of SDSS
J1536+0441 and of the surrounding galaxy cluster is yet to be clarified.
The superposition model, where the companion emits one of the two sets of
emission lines is in possible conflict with the presently available
spectroscopic data. However, the discovery of a possible nucleus in source 
B, matched with its radio detection by \citet{wrobel09}, may support this
picture. Crucial tests for this interpretation can be derived:
from UV observations of high-ionization lines \citep[as discussed 
in][]{chornock09b}, high angular resolution optical spectroscopy
and high-resolution X--ray images.

\section*{Acknowledgments}
This research has made use of the NASA/IPAC Extragalactic Database (NED)
which is operated by the Jet Propulsion Laboratory, California
Institute of Technology, under contract with the National Aeronautics
and Space Administration.

Facilities:
\facility{VLT(FORS2)}
\facility{VLT(HAWK-I)}

\label{lastpage}

\end{document}